\documentclass[twoside,11pt]{amsart}
\usepackage{amsmath}  

\newtheorem{theorem}{Theorem}

\theoremstyle{remark}

\newcommand{\new}{\newcommand} 
\new{\bg}{\begin}

\new{\iii}{\begin{enumerate}} 
\new{\fff}{\end{enumerate}}
\new{\iiii}{\begin{itemize}} 
\new{\ffff}{\end{itemize}}
\new{\mfi}{\begin{eqnarray*}} 
\new{\mff}{\end{eqnarray*}} 
\new{\mfni}{\begin{eqnarray}}
\new{\mfnf}{\end{eqnarray}}
\new{\beeq}[2]{\begin{equation}\label{#1}{#2}\end{equation}}
\new{\eqn}[1]{~(\ref{#1})}

\new{\room}{\ \ \ \ }

\new{\nor}[1]{\left\|{#1}\right\|}
\new{\scal}[2]{\left\langle{#1},{#2}\right\rangle}
\new{\set}[1]{\{{#1}\}} 
\new{\kk}{{\mathcal K}}
\new{\hh}{{\mathcal H}}
\new{\mm}{{\mathcal M}}
\new{\nn}{{\mathcal N}}
\new{\ki}[1]{{\chi}_{#1}}
\new{\runo}{{\mathbb R}}
\new{\cuno}{{\mathbb C}}
\new{\nat}{{\mathbb N}}
\new{\ccc}{\circ}
\new{\eps}{\epsilon}
\new{\supp}{\mathrm{supp\,}}

\new{\mug}{\mu_G}
\new{\muh}{\mu_H}
\new{\dgg}{\Delta_G}
\new{\deh}{\Delta_H}
\new{\intg}[2]{\int_G{#1}d\mu_G(#2)}
\new{\inth}[2]{\int_H{#1}d\mu_H(#2)}
\new{\intx}[2]{\int_X{#1}d\nu(#2)}
\new{\rap}[1]{\frac{\deh(#1)}{\dgg(#1)}}
\new{\uno}{^{-1}}
\new{\ccg}{\mathrm{C_c}{(G)}}
\new{\ccx}{\mathrm{C_c}{(X)}}
\new{\ccgg}{\mathrm{C_c}{(G\times G)}}
\new{\ind}{L^\sigma}
\new{\hind}{{\mathcal F}^\sigma}
\new{\ale}{\text{-a.e.}}
\new{\boxx}{{\mathcal B}(X)}
\new{\lind}{E^\sigma}
\new{\gdo}{\mathcal D}
\new{\ft}{\tilde{f}}

\begin{document}

\title{Square-integrability modulo a subgroup}
\author{G. Cassinelli}
\address{Gianni Cassinelli, Dipartimento di Fisica,
Universit\`a di Genova, I.N.F.N., Sezione di Genova, Via Dodecaneso~33,
16146 Genova, Italy}
\email{cassinelli@genova.infn.it}
\author{E. De Vito}
\address{Ernesto De Vito, Dipartimento di Matematica, Universit\`a di
Modena, Via Campi 213/B, 41100 Modena, Italy and I.N.F.N., Sezione di Genova,
Via Dodecaneso~33, 16146 Genova, Italy}
\email{devito@unimo.it}
\date{\today}

\begin{abstract}
A new proof of Imprimitivity theorem for transitive systems of covariance is given
and a definition of square-integrable representation modulo a subgroup is proposed.
This  clarifies the relation between coherent states, wavelet transforms and
covariant localisation observables. 
\end{abstract}

\maketitle


\section{Introduction}

In the present paper we give a self-contained proof of Imprimitivity theorem for {\em
systems of covariance}, or {\em generalised imprimitivity systems},  
based on transitive spaces. The theorem holds for locally compact groups and
non-normalised positive operator valued (POV) measures. For projective valued
measures, the theorem was proven by Mackey, \cite{mackey}, for separable
groups, and by Blattner, \cite{bla}, in full generality, and it is known as
Mackey Imprimitivity theorem.
For normalised POV measures, there are many independent proofs.  Up to our
knowledge, Poulsen, \cite{poulsen}, first proves it for Lie groups using
elliptic regularity theory, Davies, \cite{davies}, and  Scutaru,
\cite{scutaru}, for topological groups, but with some unnecessary assumption,
Neumann, \cite{neu}, and Cattaneo, \cite{cattaneo}, for locally compact 
groups. These last proofs are based on Neumark dilation theorem in order to reduce the
problem to the projective case, and on Mackey imprimitivity theorem. Finally,
Castrigiano and Henrichs, \cite{dg}, show the above result using the
theory of positive functions on a $C^*$-algebra.    

Our proof is independent both on Neumark dilation and on Mackey
Imprimitivity theorems, which are corollaries of the main result. It is based on
the proof of Mackey theorem given by Orsted, \cite{or}, as suggested by a
remark in \cite{dg} (compare also with \cite[Ch.~XXII, Sec.~3, 
Ex.~10]{dieu6}). In particular, we use a realisation of the induced
representation inspired by an exercise of \cite[Ch.~XXII, Sec.~3,
Ex.~10]{dieu6}. Our construction is a variation of the one given by
Blattner, \cite{bla}, and, in our opinion, is very elementary and {\em intrinsic},
it does not use the notion of quasi-invariant measure and the Hilbert space
where the representation acts is a space of square-integrable functions,
compare with Folland, \cite[Ch. 6]{f}.   

As a consequence of this approach, one has a {\em weak}
characterisation of the space of the intertwining operators of the induced
representation. If the group is compact, this result reduces to the Frobenius
reciprocity theorem, but, for a locally compact group, it is not completely
satisfactory. However, it clarifies the relation between  
covariant frames and systems of covariance, as suggested by many authors.
In particular, we give a definition of square-integrable representation modulo a subgroup that unifies many
notions  used in literature, for a review see \cite{alibook}, and
we obtain a characterisation of systems of covariance that extends the results
of Scutaru, \cite{scutaru}, and Holevo, \cite{h}.  

The paper is organised in the following way. In Sec.~2 we introduce the
notation and we give the construction of the induced representation. We recall
also the notion of G\"arding domain that is the main tool of our approach.
In Sec.~3 we prove the Weak Frobenius theorem and, as a consequence, we give
the definition of square-integrable representation modulo a subgroup.
In Sec.~4 we prove the Generalised Mackey theorem and, as a corollary, we 
characterise the systems of covariance.  

To avoid technical problems with integration theory, we assume that groups and
Hilbert spaces are separable, but the results hold without this hypothesis.

\section{Notations}
In this paper,  $G$ is a locally compact second countable topological group
and $H$ a closed subgroup of $G$. We denote by $\mug$ and $\muh$  left
Haar measures on $G$ and $H$, respectively. Let  $\dgg$ and $\deh$ be the
corresponding modular functions. 

Let $X=G/H$ be the quotient space of the left cosets with the natural
topology and $p:G\to X$ the canonical projection, which is an open map. 
For all $g\in G$, we denote by $x\mapsto g[x]$ the action of $G$ on $X$. If $f$ is a
function on $G$ and $g\in G$, we let $f^g$ be the map  given by
$(f^g)(g')=f(g\uno g')$, for all $g'\in G$.

Given a locally compact second countable topological space $Y$, by {\em Radon
measure} on $Y$, we mean a positive measure defined on the $\sigma$-algebra
$\mathcal{B}(Y)$ of Borel subsets of $Y$ such that it is finite on
compact sets. Since the space is second countable, Radon measures
are both outer and inner regular. In particular Haar measures are Radon.
We denote by $C_c(Y)$ the space of continuous complex functions on $Y$ with
compact support and by $\supp f$ the support of a continuous function
$f$. 

We recall the following relation between $\ccg$ and $\ccx$, due to Weil (for
the proof see, for example, Prop. 2.48 of \cite{f}).
\bg{lemma}\label{tilde}
Let $f\in\ccg$ and $K$ its support. There is a unique $\ft\in\ccx$ such that,
for all $g\in G$,
$$\ft(p(g))= \inth{f(gh)}{h}.$$
Moreover 
\beeq{con}{\sup_{x\in X}{|\ft(x)|}\leq C_K \sup_{g\in G}{|f(g)|},}
where $ C_K$ is a constant depending only on the support of
$f$. Finally, if $f'\in\ccx$ $[$positive$]$, there is $f\in\ccg$ $[$positive$]$ such
that $\ft=f'$ and $p(\supp f)=\supp f'$.
\end{lemma}

By {\em Hilbert space}, we mean a complex separable Hilbert space, being
$\scal{\cdot}{\cdot}$ the scalar product, 
linear in the first variable, and $\nor{\cdot}$ the
corresponding norm. If $A$ is a (bounded) operator, we denote by $\nor{A}$
also the norm of $A$. By a representation of $G$, we mean a
continuous (with respect to the strong operator topology) unitary
representation of $G$ acting in a Hilbert space. 
Given a representation $\pi$ acting in $\hh$, for all $f\in\ccg$, we let 
$$\pi(f)=\intg{f(g)\pi(g)}{g},$$
where the integral is in the strong operator topology. In particular,
one has that, for all $g\in G$ and $f\in\ccg$,
\beeq{0}{\pi(g)\pi(f) =\pi(f^g).}
We denote by $\gdo_\pi$ the G\"arding domain of $\pi$, {\em i.e.}, 
$$ \gdo_\pi=\mathrm{span}\set{\pi(f)u \ | \ f\in\ccg,\ u\in \hh},$$
One has the following properties.
\bg{lemma}\label{garding}
With the above notations, the G\"arding domain of $\pi$ is a $G$-invariant
dense subspace of $\hh$. If $\pi'$ is another representation of $G$ acting in $\hh'$ and $W$ is
an operator from $\hh$ to $\hh'$ intertwining $\pi$ and $\pi'$, then
$$W\gdo_\pi\subset\gdo_{\pi'}.$$  
\end{lemma}
\bg{proof}
Let $g\in G$, $f\in\ccg$ and  $v\in\hh$. By Eq.\eqn{0}, $\pi(g)\pi(f)v
=\pi(f^g)v$ and,  since $f^g\in\ccg$, it follows that $\gdo_\pi$ is
$G$-invariant. To show the density, given $v\in\hh$
and $\eps>0$, since $\pi$ is continuous with respect to the strong operator
topology, there is a compact neighbourhood $K$ of the identity such that, 
for all $g\in K$, $\nor{\pi(g)u -u}\leq \eps$. Since $K$ contains a non-void
open set, $\mug(K)>0$ and, by outer regularity of $\mug$, there is an open set $V\supset K$
with $\mug(V\backslash K)\leq\eps \mug(K)$.
Let $f\in\ccg$ such that $f(g)=1$ for all $g\in K$, $0\leq
f(g)\leq 1$ for all $g\in G$ and supp$f\subset V$. Then, defined  $a=\frac{1}{\mug(K)}$,
\mfi
\nor{\pi(af)v -v} & = & \nor{a\int_K(\pi(g)v - v)d\mug(g) + a
\int_{V\backslash K} f(g)\pi(g)v d\mug(g)} \\
 & \leq & a\int_K \nor{\pi(g)v - v}d\mug(g) + a
\int_{V\backslash K} f(g)\nor{\pi(g)v} d\mug(g) \\
& \leq &  \eps (1+  \nor{v}) .
\mff 

To show the second point, let $f\in\ccg$ and $v\in\hh$, then 
\mfi
W\pi(f)v & = & W\intg{ f(g)\pi(g)v}{g} = \\ 
\intg{ f(g)W\pi(g)v}{g} & = & \intg{ f(g)\pi'(g)Wv}{g} =  \pi'(f)Wv.
\mff
\end{proof}
If $G$ is a Lie group,usually G\"arding domain is defined replacing $\ccg$
with $C^{\infty}_c(G)$, see \cite{warner}. We adopted the definition of \cite{or}. 

We now give a realisation of the induced representation, based on the
following lemma. 
\bg{lemma}\label{theta}
There is a continuous function $\theta: G \to [0,+\infty[$ such that, for all $g\in G$, 
$$\inth{\theta(gh)}{h}= 1,$$
and, for any compact subset $K$ of $G$,  $KH\cap\supp \theta$ is compact.

Moreover, let $Y\in\mathcal{B}(G)$ such that, for all $h\in H$,
$\mug(Yh\backslash Y) =0$. Then $Y$ is
negligible with respect to $\theta\mug$ if and only if negligible with respect
to $\mug$, where $\theta\mug$ is the measure having density $\theta$ with
respect to $\mug$. 
\end{lemma}
\bg{proof}
The existence of $\theta$ is proven, for example, in Prop.~2 of
\cite{gaal}. With respect to second part, if $\mug(Y)=0$, then
$\int_Y\theta(g)d\mug(g)=0$. Conversely,
\mfi
\int_Y\theta(g)d\mug(g)  & = & \int_Y\theta(g)\inth{\theta(gh)}{h}d\mug(g) \\
& = & \inth{ \int_Y\theta(gh)d\mug(g)}{h} \\
(g\mapsto gh\uno) & = & \inth{\dgg(h\uno) \int_{Yh}\theta(g)d\mug(g)}{h} \\
& = & \inth{\dgg(h\uno) \int_{Y}\theta(g)d\mug(g)}{h} \\
& = & 0,
\mff
where we used that $\mug(Yh\backslash Y)=0$.
\end{proof}

Let $\sigma$  be a representation  of $H$ acting in $\kk$. Given $\theta$ as in
the above lemma, let $\hind$ be the subspace of functions $F$ from $G$
to $\kk$ such that
\iiii
\item $F$ is $\mug$-measurable;
\item given $h\in H$, for $\mug$-almost all $g\in G$, 
\beeq{covar}{\sqrt{\rap{h}}\sigma(h\uno)F(g)=F(gh);}
\item $\intg{\nor{F(g)}^2\theta(g)}{g} <+\infty$.
\ffff  
We notice that, due to Lemma \ref{theta}, a function $F$ satisfying Eq.\eqn{covar} 
is $\mug$-measurable if and only if it is $\theta\mug$-measurable, so
$F\in\hind$ if and only if $F\in L^2(G,\theta\mu_G,K)$ and
Eq.\eqn{covar} holds. 

Given $v\in\kk$ and $f\in\ccg$, let $F_{f,v}$ be the
function from $G$ to $\kk$ defined, for all $g\in G$, as
$$(F_{f,v})(g) = \inth{\sqrt{\rap{h\uno}}f(gh)\sigma(h)v\ }{h}.$$

\bg{lemma}\label{fv}
With the above notations, the space $\hind$ is a closed subspace of
$L^2(G,\theta\mu_G,K)$, which does not depend on the choice of
$\theta$, and  each $F\in\hind$ is locally $\mug$-integrable. For each 
$f\in\ccg$ and $v\in\kk$,  $F_{f,v}$ is in $\hind$, it is continuous and
$\supp F_{f,v}\subset (\supp f)H$.  Finally, the space generated by the elements
of the form  $F_{f,v}$ is dense in $\hind$.
\end{lemma} 
\bg{proof}
We claim that $\hind$ is closed.  Indeed, let $(F_n)_{n\in \nat}$ be a
Cauchy sequence in $\hind$. Since $L^2(G,\theta\mu_G,K)$ 
is a Hilbert space, 
$(F_n)_{n\in \nat}$ converges in $L^2$ and, possibly passing to a
subsequence, $\theta\mug$-almost everywhere. Let $Y$ be the complement of the set of
elements $g\in G$ such that $(F_n(g))_{n\in\nat}$ converges
pointwise and denote by $F(g)$ the limit. By hypothesis, $Y$ is
$\theta\mug$-negligible and, by unicity of the limit, $(F_n)_{n\in \nat}$ converges to $F$  in $L^2$.
Let now $h\in H$, by definition of $\hind$
and the fact that $(F_n)_n$ is denumerable, it exists $Y_h\in\mathcal{B}(G)$ such
that $\mug(Y_h)=0$ and, for all $g\in G\backslash Y_h$ and $n\in\nat$
$$ \sqrt{\rap{h}}\sigma(h\uno)F_n(g)=F_n(gh).$$
If $g\not\in Y\cup Y_h$, passing to the limit, one has that $(F_n(gh))_{n\in\nat}$ converges
and
\beeq{p}{\sqrt{\rap{h}}\sigma(h\uno)F(g)=F(gh).}
In particular, $gh\not\in Y$, that is $Yh\uno \subset  Y\cup Y_h$. 
Since $\mug(Y_h)=0$, it follows that $\mug(Y_h\backslash Y)=0$. By
Lemma~\ref{theta}, it follows that $\mug(Y)=0$ and, hence, 
$\mug(Y\cup Y_h)=0$.  So Eq.\eqn{p} holds
$\mug$-almost everywhere, that is $F\in\hind$.

We now prove that $\hind$ is independent on $\theta$. Let $\theta'$ a
non-negative continuous function such that $\inth{\theta'(gh)}{h}=1$ for
all $g\in G$. Let $F$ from $G$ to $K$ $\mug$-measurable and such that
Eq.\eqn{covar} holds. Then
\mfi
\intg{\nor{F(g)}^2\theta(g)}{g} & = & \intg{\nor{F(g)}^2\theta(g)\inth{\theta'(gh)}{h}}{g} \\
& = & \inth{ \intg{\nor{F(g)}^2\theta(g)\theta'(gh)}{g}}{h} \\
(g\mapsto gh\uno) & = & \inth{\intg{\nor{F(gh\uno)}^2\theta(gh\uno)\theta'(g)\dgg(h\uno)}{g}}{h} \\
(F\in\hind, h\mapsto h\uno) & = & \inth{\intg{\nor{F(g)}^2\theta(gh)\theta'(g)}{g}}{h} \\
& = & \intg{\nor{F(g)}^2\theta'(g)\inth{\theta(gh)}{h}}{g} \\
& = &\intg{\nor{F(g)}^2\theta'(g)}{g}.
\mff
This shows the claim.

Let now $F\in\hind$, we prove that $F$ is locally
$\mug$-integrable. Let $f\in\ccg$ non-negative, then, as before,
\mfi
\intg{\nor{F(g)} f(g)}{g} & = & \intg{\nor{F(g)}f(g)\inth{\theta(gh)}{h}}{g} \\
& = &   \intg{\nor{F(g)}\theta(g)\inth{\sqrt{\rap{h\uno}} f(gh)}{h}}{g} \\
& = & \intg{\nor{F(g)}f'(g)\theta(g)}{g} 
\mff
where $f'(g):= \inth{\sqrt{\rap{h\uno}} f(gh)}{h}$ is continuous and
$\supp f'\subset(\supp f)H$. By Lemma \ref{theta}, $f'\theta\in L^2(G,\mug)$,
so that $\nor{F(g)}f'(g)\theta(g)$ is $\mug$-integrable, hence $F$ is locally
$\mug$-integrable.

The properties of $F_{f,v}\in\hind$ are clear (use the proof
of Lemma \ref{tilde}). We show the density. Let $F\in\hind$ such that, for all
$f\in\ccg$ and $v\in\kk$, $ \scal{F}{F_{f,v}}=0$. Then, using the same argument as
before and Tonelli theorem, one can check that the map
$$(g,h)\mapsto
\scal{F(g)}{\sigma(h)v}\overline{f(gh)}\theta(g)\sqrt{\rap{h\uno}} $$
is $\mug\otimes\muh$-integrable and 
\mfi
0 & = &  \scal{F}{F_{f,v}} \\
& = &
\intg{\inth{\scal{F(g)}{\sigma(h)v}\overline{f(gh)}\theta(g)\sqrt{\rap{h\uno}}}{h}}{g} \\
(g\mapsto gh\uno, h\mapsto h\uno)) 
& = & \intg{\scal{F(g)}{v}\overline{f(g)}\inth{\theta(gh)}{h}}{g} \\
& = & \intg{\scal{F(g)}{v}\overline{f(g)}}{g}.
\mff
By standard arguments, one has that $F(g)=0$ $\mug$-almost all $g\in
G$, that is $F=0$.
\end{proof} 
Define, for all $g\in G$ and $F\in\hind$,
$$ (\ind_g F)(g')=F(g\uno g')\room g'\in G,\ \mu_G\ale\ .$$ 
One has the following result.
\bg{proposition}\label{ind}
Let $\sigma$ be a representation of $H$, then $\ind$ is a
representation of $G$ acting in $\hind$ and is a realisation of the
representation induced  by $\sigma$ from $H$ to $G$.
In particular, the G\"arding domain of $\ind$ is a subspace of continuous functions.
\end{proposition}

\bg{proof}
Given $g\in G$, we prove that $\ind_g$ is a well-defined isometric
operator in $\hind$. Let
$F\in\hind$, then, for all $h\in H$ and for $\mug$-almost all $g'\in G$
$$(\ind_g F)(g'h)=F(g\uno g'h)= \sqrt{\rap{h}}\sigma(h\uno)F(g\uno g')= \sqrt{\rap{h}}\sigma(h\uno)(\ind_g F)(g').$$
Moreover,
\mfi
\intg{\nor{F(g\uno g')}^2 \theta(g')}{g'} & = &  \intg{\nor{F(g')}^2\theta(gg')}{g'} \\
& = &  \intg{\nor{F(g')}^2\theta(gg')\inth{\theta(g'h)}{h}}{g'} \\
& = &  \inth{\intg{\nor{F(g')}^2\theta(gg')\theta(g'h)}{g'}}{h} \\
(g'\mapsto g'h\uno) & = & \inth{\intg{\dgg(h\uno)\nor{F(g')}^2\theta(gg'h\uno)\theta(g')}{g'}}{h} \\
(h\mapsto h\uno) & = & \intg{\nor{F(g')}^2\theta(g')  \inth{\theta(gg'h)}{h}}{g'} \\
& = & \intg{\nor{F(g')}^2\theta(g') }{g'} 
\mff
This proves that $\ind_g$ is a well defined isometric operator in $\hind$. 

In order to show that $g\mapsto \ind_g$ is continuous, since  $\ind_g$ is
isometric and by a density argument, one can reduced to prove that,
given $f\in\ccg$, $v\in\kk$ and $F'\in\hind$, the map 
$$g\mapsto \scal{\ind_g F_{f,v}}{F'}=\intg{\scal{F_{f,v}(g\uno g')}{F'(g)}\theta(g)}{g}$$
is continuous. However, due to Lemma~\ref{fv}, $F_{f,v}$ is continuous
and, due to Lemma \ref{theta}, $\supp F_{f,v}\cap \supp
\theta$ is compact, so that the thesis follows by dominated convergence theorem. 
  
We prove that $\gdo_\ind$ is a subspace of continuous functions. Indeed, let $f\in\ccg$ and $F\in\hind$. Given $F'\in\hind$,
observe that the function on $G\times G$ 
$$\Psi(g,g')= f(g) \scal{F(g\uno g')}{F'(g')}\theta(g')$$
is $\mu_G\otimes\mu_G$-measurable and 
\mfi
\intg{\intg{| f(g) \scal{F(g\uno g')}{F'(g')}\theta(g') | }{g'}}{g} & \leq  \\ 
\intg{\intg{ |f(g)| \nor{F(g\uno g')} \nor{F'(g')}\theta(g') }{g'}}{g} & \leq \\
\intg {|f(g)|\left(\intg{ \nor{F(g\uno g')}^2\theta(g')}{g'}\right)^{\frac
    12}\left(\intg{ \nor{F'(g')}^2\theta(g')}{g'}\right)^{\frac 12}}{g}  & = \\
  \intg{ |f(g)| \nor{\ind_g F}_{\hind}  \nor{F'}_{\hind}}{g} & = \\  
\intg{ |f(g)| \nor{F}_{\hind}  \nor{F'}_{\hind}}{g}& .
\mff
Since $f$ has compact support, the above integral is finite and, by Tonelli theorem,
$\Psi$ is integrable with respect to $\mug\otimes\mug$. Then
\mfi
\scal{\ind(f)F}{F'} & = & \intg{ f(g)\intg{ \scal{F(g\uno g')}{F'(g')}\theta(g')}{g'}}{g} \\
& = &  \intg{\intg{ \scal {f(g) F(g\uno g')}{F'(g')}}{g}\theta(g')}{g'} \\
(g\mapsto g'g,g\mapsto g\uno)& = &   \intg{\intg{ \scal {\dgg(g\uno) f(g'g\uno) F(g)}{F'(g')}}{g}\theta(g')}{g'} \\
& = & \intg{\scal{(f\star  F)(g')}{F'(g')}\theta(g')}{g'},
\mff
where $(f\star F)(g') =\intg{\dgg(g\uno) f(g'g\uno) F(g)}{g}$, which is well defined since $F$ is locally $\mu_G$-integrable.
Then, one has that $\ind(f)F= f\star F$. The continuity of $f\star F$
is now consequence of the dominated convergence theorem.
\end{proof}

\section{Weak Frobenius theorem}

The following definition is a possible extension of the notion of admissible
vector for square-integrable representations. We fix a function
$\theta$ as given by Lemma~\ref{theta}. 
\bg{definition}
Let  $\sigma$ be a representation of $H$ acting in $\kk$ and $\pi$ a representation of $G$ acting in
$\hh$. A linear map  $A:\gdo_\pi \to\kk$ such that
\bg{itemize}
\item for all $h\in H$ and $v\in\gdo_\pi$,
\beeq{semi}{\sigma(h) A v= \sqrt{\rap{h}} A\pi(h) v;} 
\item for all $v\in\gdo_\pi$, the map from $G$ to $\kk$
$$g\mapsto A\pi(g\uno)v:=(W_Av)(g)$$ 
is $\mug$-measurable and   
\beeq{square}{\intg{ \nor{A\pi(g\uno)v}^2 \theta(g)}{g}\leq \beta  \nor{v}^2,}
where $\beta$ is a positive constant independent on $v$,
\end{itemize}
is called {\em admissible map for $\pi$ modulo $(H,\sigma)$}. 
\end{definition}
The admissible maps modulo $(H,\sigma)$ give a characterisation of the commuting ring of the
representation induced by $\sigma$, compare with the results obtained by
Moore, \cite{moore}.
\bg{theorem}[Weak Frobenius theorem]
Let  $\sigma$ be a representation of $H$ acting in $\kk$ and $\pi$ a representation of $G$ acting in
$\hh$. Let $A:\gdo_\pi \to\kk$ be an  admissible map for $\pi$ modulo $(H,\sigma)$, then 
\iiii
\item for all $v\in\gdo_\pi$, $W_Av\in{\gdo_{\ind}}\subset\hind$ (in particular
  $W_Av$ is a continuous function); 
\item the linear map $v\mapsto W_Av$ extends to a unique bounded operator $W_A$,
  called {\em wavelet transform}, from $\hh$ to $\hind$ that intertwines $\pi$ and $\hind$.
\ffff
Conversely, given a bounded operator $W:\hh\to\hind$ intertwining $\pi$ with
$\hind$, there is a unique admissible map $A$ (for $\pi$ modulo
$(H,\sigma)$) such that,  for all $v\in\gdo_\pi$, $Wv=W_A v$.
\end{theorem}

\bg{proof}
Let $A$ be an admissible map. Given $v\in\gdo_\pi $ and $h\in H$,
\mfi
(W_Av)(gh) & = & A \pi(h\uno g\uno )v = \\
\sqrt{\rap{h}}\sigma(h\uno) A\pi(g\uno) v & = &
\sqrt{\rap{h}}\sigma(h\uno) (W_Av)(g),
\mff
for all $g\in G$. Due to Eq.\eqn{square}, one has that
$W_Av\in\hind$ and $\nor{W_Av}\leq\sqrt{\beta} \nor{v}$.  Since $\gdo_\pi$ is
dense in $\hh$, $v\mapsto W_Av$  extends to a unique bounded operator $W_A$. Moreover, if now $g'\in G$
$$(W_Av)({g'}\uno g) = A \pi(g\uno g')v = (W_A \pi(g')v)(g)$$
for all $g\in G$, so that $W_A$ intertwines $\pi$ and $\ind$. In particular,
due to Lemma~\ref{garding}, $W_A\gdo_\pi\subset{\gdo_{\ind}}$, 
and the elements of $\gdo_{\ind}$ are continuous functions by Prop.~\ref{ind}.

Conversely, let $W$ be bounded operator from $\hh$ to $\hind$ intertwining
$\pi$ and $\hind$. By  Lemma~\ref{garding} and Prop.~\ref{ind}, for all
$v\in\gdo_\pi$, $Wv$ is a continuous function and we can define $A$ from
$\gdo_\pi$ to $\kk$ as $Av= (Wv)(e)$, where $e$ is the identity of $G$.
Given $h\in H$ and $v\in\gdo_\pi$, 
$$ A\pi_h v =   (W\pi_h v)(e) = (\ind_h Wv)(e) = (Wv)(h\uno) = \sqrt{\rap{h\uno}}\sigma(h) (Wv)(e),$$
so that Eq.\eqn{semi} holds. Moreover, if $g\in G$
$$ A\pi_{g\uno}v = (W\pi_{g\uno} v)(e) =  (\ind_{g\uno} Wv)(e)= (Wv)(g),$$
so that $W_A =W$ on $\gdo_\pi$ and Eq.\eqn{square} is satisfied with
$\beta=\nor{W}^2$. Let $B$ an other admissible map such that, for all $v\in\gdo_\pi$
$$ B\pi_{g\uno}v = (Wv)(g)\room g\in G\ \theta\mug\ale.$$
Since both side satisfy Eq.\eqn{covar}, by Lemma~\ref{theta},
the equality holds $\mug$-almost everywhere and, by continuity, everywhere.
With the choice $g=e$, one has $Bv=(Wv)(e)=A$.
\end{proof}
We add some comments. If a linear map $A$ satisfies Eq.\eqn{semi} and is
closable, its closure is semi-invariant with weight $h\mapsto\sqrt{\rap{h}}$
in the sense of \cite{dm} and the measurability of $W_Av$ follows from the
continuity of $\pi$. However, as shown by Example ~\ref{no} below, there
are admissible maps that are not closable. If $A$ is closable and
$\pi$ is irreducible, the condition~\eqn{square} is equivalent to the
existence of  $v\in\gdo_\pi$ such that
$$0<\intg{ \nor{A\pi(g\uno)v}^2 \theta(g)}{g}<+\infty,$$
(use the first part of the proof of Th.~3 of \cite{dm}).

If $X$ admits an invariant measure (in particular if both $G$ and $H$ are
unimodular), Eq.\eqn{semi} is the requirement that $A$ is an
(algebraic) intertwining map between $\pi_{|H}$ and $\sigma$.

Assume now that $G$ is compact and $\pi$ irreducible. Since
$\hh$ is finite dimensional, $\gdo_\pi=\hh$ and, taking into account that 
both $G$ and $H$ are unimodular, Eq.\eqn{semi} is the condition that $A$ is an
intertwining operator between $\pi_{|H}$ and $\sigma$. Finally, since the
measure $\mug$ is bounded, Eq.\eqn{square} is trivially satisfied. Then, the
space of admissible maps is precisely the set of intertwining operators between
$\pi_{|H}$ and $\sigma$ and the above theorem reduced to Frobenius reciprocity
theorem for compact groups, due to Weil, see, for example, \cite{f}.

In case that $G$ is not compact, the following example shows that it is
restrictive to assume that admissible maps are closable. 
\bg{example}\label{no} 
In the above theorem, let $\pi=\ind$ and choose $W=I$. A simple computation
shows that the admissible map $A$ such that $W_A=I$
is, for all $F\in\gdo_{\ind}$,
$$AF=F(e),$$
which is clearly not closable (if $G$ is not discrete).

In particular, let $G$ be the Poincar\'e group $\runo^4\times'SO(3,1)$, 
$H=\runo^4\times SO(3)$ and $\sigma$ the trivial representation of $H$. It is well
known that $G/H$ has an invariant measure and $\ind$ is irreducible, so that
the multiples of identity are the only intertwining operators. Due to the
previous observation, in this example there are neither bounded nor closable admissible maps.  
\end{example}
Due to the fact that, in general,  G\"arding domains do not have a natural
topology such that Eq.\eqn{square} is equivalent to the continuity of the
admissible maps,  our result is not good enough to give a {\em useful}
characterisation of the set of intertwining operators between $\pi$ and
$\ind$,  compare with so-called "intertwining number theorems", see, for an exposition,
\cite{warner}, and the results contained in \cite{poulsen1} for  Lie groups.
However, the above theorem allows to characterise completely
the representations that are equivalent to a sub-representation of the induced one.
\bg{corollary}\label{def}
With the notations of the above theorem, the following conditions are equivalent: 
\iiii
\item the representation $\pi$ is equivalent to a sub-representation of $\ind$;
\item there is an admissible map $A_0$ such that, for all $v\in\gdo_\pi$,
\beeq{tight}{\intg{ \nor{A\pi(g\uno)v}^2 \theta(g)}{g}= \nor{v}^2.}
\item there is an admissible map $A$ such that, for all $v\in\gdo_\pi$,
\beeq{frame}{\alpha \nor{v}^2\leq \intg{ \nor{A\pi(g\uno)v}^2 \theta(g)}{g}\leq
  \beta  \nor{v}^2,}
where $0<\alpha\leq\beta;$
\ffff
If any of the above three conditions is satisfied, we said to $\pi$ is {\em square-integrable modulo} $(H,\sigma)$. 
\end{corollary}
\bg{proof}
Assume  first condition, then there is an isometry $W$ intertwining $\pi$ and
$\ind$. Applying Weak Frobenius theorem to $W$, there exists an admissible map $A$ such
that $W=W_A$ and, since $W$ is isometric, Eq.\eqn{tight} holds. Clearly
Eq.\eqn{tight} implies Eq.\eqn{frame}. Assume now the third 
condition, the corresponding wavelet operator $W_A$ satisfies, for all $v\in\hh$
$$\sqrt{\alpha} \nor{v} \leq \nor{W_Av} \leq \sqrt{\beta}  \nor{v}.$$ 
In particular, $W_A$ is injective, so that, by polar decomposition, there is
an isometry $W_0$ such that $W_A=W_0 |W_A|$. Since $W_A$ commutes with the
action of $G$, $W_0$ intertwines $\pi$ and $\ind$.
\end{proof}
In the framework of wavelet analysis, Eq.\eqn{frame} says that
$\set{A\pi(g\uno)}_{g\in G}$ is a (vector valued) frame in $\hh$ and 
Eq.\eqn{tight} that this frame is tight. So one can restate the above
corollary in the following way. A representation $\pi$ is square-integrable
modulo $H$ if and only if the set $\set{A\pi(g\uno)}_{g\in G}$ is a frame for
some admissible map $A$, and $A$ can always be chosen in such a way that the
corresponding frame is tight.  

\bg{example} Assume that $\pi$ is irreducible and let $H=\set{e}$ being
$\sigma$ the trivial representation. Then, $\pi$ is square-integrable modulo
$(H,\sigma)$ if and only if $\pi$ is square-integrable in the
sense of Godement, see, for example,~\cite{gaal}, if $G$ is unimodular,
and~\cite{dm}, if $G$ is non-unimodular. In particular, there exist always bounded
admissible maps $A=\scal{\cdot}{v}$ where $v$ is in the domain of the
formal degree of $\pi$, \cite{dm}, such that Eq.\eqn{tight} holds (compare
with Example~\ref{no} above and Example~\ref{fu} below).
\end{example}
The following result gives some informations when the admissible map is bounded.
\bg{corollary}
With the notations of the above theorem, let $A:\hh\to\kk$ be a bounded
operator satisfying Eq.$\eqn{semi}$. Then, 
\iii
\item the space $X$ has an invariant measure, {\em i.e.} $\rap{h}=1$ for all
  $h\in H$; 
\item if $A$ satisfies Eq.$\eqn{square}$, the corresponding wavelet operator $W_A$ is given by
$$(W_A v)(g)=A\pi(g\uno)v,$$ 
for all $g\in G$ and $v\in \hh$;
\item if $A$ satisfies Eq.$\eqn{frame}$, then $\pi$ is square-integrable
  modulo both $(H,\sigma)$ and $(H,\pi_{|H})$. 
\fff
\end{corollary}
\bg{proof}
With respect to the first claim, it is clear that, if $A$ satisfies Eq.\eqn{semi}
for all $v\in\gdo_\pi$, then  Eq.\eqn{semi} holds for all $v\in\hh$, 
{\em i.e.} $A$ is is semi-invariant with weight $h\mapsto\sqrt{\rap{h}}$ in the sense of
\cite{dm}. However, $A$ is bounded and this is possible only if 
$\rap{h}=1$ for all $h\in H$, compare with \cite[Eq.~2]{dm}.

In order to show the second statement, let $v$ in $\hh$ and $(v_n)_{n\in\nat}$ in
$\gdo_\pi$ such that $v=\lim v_n$. Since $\left((W_Av_n)(g)\right)_n$ converges pointwisely
to the continuous function $\psi_v(g)=A\pi(g\uno)v$,  by
Eq.\eqn{square} and Fatou lemma, it follows that $\psi_v\in\hind$. On the
other hand, $W_Av=\lim W_Av_n$ and, by unicity of the limit, $\psi_v=W_Av$. 

Finally assume that Eq.\eqn{frame} holds and let $A=U|A|$ be the polar decomposition of
$A$. Clearly $|A|$ commutes with $\pi_{|H}$ and, taking into account that $U$ restricted to
the range of $|A|$ is an isometry, Eq.\eqn{frame} becomes
$$\alpha \nor{v}^2\leq \intg{ \nor{|A|\pi(g\uno)v}^2
\theta(g)}{g}\leq \beta  \nor{v}^2, $$
so, by the above corollary $\pi$ is square-integrable modulo $(H,\pi_{|H})$.
\end{proof}
From the above corollary it follows that if $\pi$ is square-integrable modulo
$(H,\sigma)$ with respect to some bounded admissible map, then it is
square-integrable modulo $(H,\pi_{|H})$.  However, also with this assumption, in general
Eq.\eqn{tight}  can not be satisfied by any bounded admissible map, as showed
by the following example, adapted from \cite{fuhr1} (compare with the notion of weak and strong
square-integrability in \cite{fuhr1}).
\bg{example}\label{fu} 
Let $G=\runo$ and $\pi$ be the left regular representation,
$H=\set{e}$ and $\sigma$ the trivial representation.  Clearly, any bounded
admissible map is of the form $\scal{\cdot}{v}$ for some vector $v\in  L^2(\runo)$.
Losert and Rindler, \cite{lr}, prove  that there is a vector $\eta\in L^2(\runo)$ with compact support and cyclic. 
Let $A=\scal{\cdot}{\eta}$, then  the corresponding operator $W_A$ is injective so
that $\pi$ is in fact square-integrable (modulo $(H,\sigma)$). However, since
$\runo$ is unimodular and not discrete, F\"uhr and Mayer, \cite{fuhr1}, show that
there are not vectors $v\in\hh$ such that $A_0=\scal{\cdot}{v}$ satisfies Eq.\eqn{tight}. 
\end{example}
Our definition of square-integrability modulo a subgroup unifies many
notions  used in literature in the fields of wavelet analysis
and of generalised coherent states. For example.
\iii
\item Square-integrability modulo the centre: $\pi$ is irreducible, $H$ is a central subgroup
of $G$, $\sigma$ is the character of $H$ defined by the restriction of $\pi$ to
$H$ and $A=\scal{\cdot}{v}$, for some non-zero vector $v\in\hh$, \cite{borel}; 
\item Gilmore-Perelemov coherent states and $\alpha$-admissible vectors: $\pi$ is cyclic, $H$ is the stability subgroup, up to a phase
  factor, of some non-zero vector $v\in\hh$ with respect to the action of $\pi_{|H}$, 
  $\sigma$ is the corresponding character of $H$ and $A=\scal{\cdot}{v}$, \cite{alibook}, \cite{perelemov},
  \cite{s} and reference therein.
\item Systems of coherent states: $\pi$ is arbitrary, $\sigma=\pi_{|H}$ and
  $A^*A$ is of trace class, \cite{klauder}, \cite{scutaru}.
\item Vector coherent states and $V$-admissible vectors: $\pi$
  is arbitrary, $\sigma$ is a finite dimensional representation
 contained in the restriction of $\pi$ to $H$ and $A$ is the projection on the
 closed subspace left invariant by $\sigma$, \cite{ali}, \cite{alibook}, \cite{rowe} and reference therein. 
\item Weak and strong integrability: $\pi$ is arbitrary, $H=\set{e}$ with the
  trivial representation, and $A=\scal{\cdot}{v}$ for some $v\in\hh$,
  \cite{ali1}, \cite{fuhr} and \cite{fuhr1}. 
\fff

\section{Generalised Imprimitivity theorem}
We start with the definition of covariant localisation observable.
\bg{definition}
Given a representation $\pi$ of $G$ acting in $\hh$, a map $E$ from the Borel
subsets $\boxx$ of $X$ into the set of positive operators in $\hh$ such that
\iii
\item $E(\emptyset)=0$;
\item $E(X)$ is injective;
\item for any disjoint sequence $(Y_n)_{n\in\nat}$ in $\boxx$,
$$E(\cup_n Y_n)=\sum_n E(Y_n),$$
where the series converges in the strong operator topology;
\item for all $g\in G$ and $Y\in\boxx$,
\beeq{impri}{\pi_gE(Y)\pi_g\uno = E(g[Y]),}
\fff
is called a {\em localisation observable}  based on $X$, covariant with
respect to $\pi$ and acting in $\hh$. Moreover, 
\bg{itemize}
\item if $E(X)=I$, $E$ is said to be {\em normalised}, 
\item if, for all $Y\in\boxx$, $E(Y)$ is a projection operator, $E$ is said to be {\em
projective}.
\end{itemize}
\end{definition}
The first and third requirement is the fact that $E$ is a
POV measure on $X$ and the forth that $(\pi,E)$ is a {\em
system of $G$-covariance}, \cite{cattaneo}, or a {\em generalised imprimitivity}, 
\cite{h} (see, also, \cite{davies}, \cite{poulsen}). The second requirement is not a
constraint, since the kernel of $E(X)$ is invariant with respect to
the action of $\pi$ and is contained by the kernel of $E(Y)$ for any
$Y\in\boxx$. Finally, if $E$ is projective, then it is
necessarily normalised and commutative and  $(\pi,E)$ is a system of
imprimitivity, \cite{mackey} (see, also, \cite{f}).
The reason to introduce the name {\em covariant localisation observable},
instead of  {\em system of $G$-covariance}, is to stress the different role
between the representation $\pi$ and the POV measure $E$. In doing so, we adopt the
terminology from  Quantum Mechanics, see, for example, \cite{pekka}, \cite{holevo}. 

The notion of equivalence is the natural one. Indeed, if $E_1$ and $E_2$ are
localisation observables  covariant with respect to 
$\pi_1$ and $\pi_2$, respectively, they are  {\em equivalent} if there is a
unitary operator $T$ intertwining $\pi_1$ and $\pi_2$ such that
$$E_2(Y)T=TE_1(Y)\room \forall Y\in\boxx.$$

\bg{example}\label{ex} 
Let $\sigma$ be a representation of $H$
and $\ind$ the corresponding induced representation acting in $\hind$.
For all $Y\in\boxx$, let $\lind(Y)$ be the operator in $\hind$ defined by 
$$(\lind(Y)F)(g)=\chi_{Y}(p(g)) F(g)\room g\in G,\ \mug\ale,$$
where $F\in\hind$ and $\chi_Y$ is the characteristic function of the subset $Y$. 
It is well known, see, for example, \cite{f}, that $Y\mapsto E(Y)$ is a
projective localisation observable  based on $X$ and  covariant with
respect to $\ind$. 

Let now $T$ be a positive operator in $\hind$ commuting with
$\ind$. Define $\hind_T$ as the closure of the range of $T$, $\ind_T$ be the
restriction of $\ind$ to $\hind_T$  and, for any $Y\in\boxx$,
$\lind(Y)_T=T\lind(Y)T$, regarded as operator in $\hind_T$. Clearly,
the map $Y\mapsto\lind(Y)_T$ is localisation observable  based on $X$ and  covariant with
respect to $\ind_T$. The next theorem will show that, up to an
equivalence, all the localisation observables are of this form.
\end{example}    

\begin{theorem}[Generalised Mackey theorem] 
Let $\pi$ be a representation of $G$ acting in $\hh$ and $E$  a localisation observable based
on $X$ covariant with respect to $\pi$. There is a  unique (up to
an equivalence class) representation $\sigma_E$ of $H$ and an isometry $W$ from
$\hh$ to ${\mathcal F}^{\sigma_E}$ such that 
\mfni
W \pi(g)  & = &  L^{{\sigma}_E}(g) W\room\room \room\room g\in G \label{uno}\\
E(Y) & = &  E(X)^{\frac 12} W^* E^{\sigma_E}(Y) W E(X)^{\frac 12}\room\room  Y\in\boxx\label{due}\\
\hind& =& \overline{\mathrm{span}}\set{E^{\sigma_E}(Y)W v \ | \ Y\in\boxx, v\in\hh}.\label{tre}
\mfnf
Moreover, $E$ is projective if and only if $WE(X)^{\frac 12}$
unitary. Finally, if $E'$ is another localisation observable equivalent to $E$, 
then $\sigma_{E'}$ is equivalent to $\sigma_E$.
\end{theorem}

\bg{proof} We split the proof in seven steps.

\noindent{\em Step }1). We define an operator valued linear form $M$ on $\ccg$
associated with the POV measure $E$. 

Given $u\in\hh$, let $d\scal{E(x)u}{u}$ be the bounded (Radon) measure on $X$
$$Y\mapsto \scal{E(Y)u}{u},$$
having total mass $ \scal{E(X)u}{u}\leq \nor{u}^2 \nor{E(X)}$ and satisfying,
due to Eq.\eqn{impri}, 
\beeq{30}{\int_X f(g[x]) d\scal{E(x)u}{u} =  \int_X f(x) d\scal{E(x)\pi_g u}{\pi_g u},}
for all $g\in G$. 

Given $f\in\ccg$ and $u\in\hh$, by Lemma~\ref{tilde}, the integral
$$\int_{X} \left(\inth{f(gh)}{h}\right) d\scal{E(p(g))u}{u},$$
is well defined, linear in $f$, quadratic in $u$  and it is bounded by
$$ C_f \sup_{g\in G}{|f(g)|} \nor{u}^2 \nor{E(X)},$$
where $C_f$ is a constant depending only on the support on $f$.
Fixed $f\in\ccg$, by polarisation identity, there is a unique operator $M(f)$
on $\hh$ such that, for all $u\in\hh$,
\mfni
\scal{M(f)u}{u} & = & \int_{X} \left(\inth{f(gh)}{h}\right)d\scal{E(p(g))u}{u} \nonumber\\
\nor{M(f)} & \leq & 4 C_f  \sup_{g\in G}{|f(g)|} \nor{E(X)},\label{19}
\mfnf
(the factor 4 is due to polarisation identity). By Eq.\eqn{30}, it follows that,
for all $g\in G$ and $f\in\ccg$,
\beeq{2}{\pi(g)M(f)\pi(g)=M(f^g).}
We claim that, for all $h\in H$ and $f\in\ccg$, 
\beeq{6}{M(f(\cdot\,h))=\deh(h\uno)M(f),}
where $f(\cdot\,h)$ is the function $g\mapsto f(gh)$. 
Indeed, if $u\in\hh$,
\mfi 
\scal{M(f(\cdot\,h))u}{u} & = & \int_{X} \left(\inth{f(gsh)}{s}\right)
d\scal{E(p(g))u}{u},\\
& = & \int_{X} \left(\inth{f(gsh)}{s}\right) d\scal{E(p(g))u}{u} \\
(s\mapsto sh\uno)& = & \deh(h\uno) \int_{X} \left(\inth{f(gs)}{s}\right) d\scal{E(p(g))u}{u} \\
& = & \deh(h\uno) \scal{M(f)u}{u}.
\mff
\vspace{.3cm}

\noindent{\em Step }2). We show that, if $u,v\in\gdo_\pi$,  there is
a unique continuous function $\phi_{u,v}$ defined on $G$ such that
$$\scal{M(f)u}{v} = \intg{f(g)\phi_{u,v}(g)}{g}\room f\in\ccg.$$

The unicity is clear, since $\mug$ is a Radon measure and $\phi_{u,v}$ is
continuous. To prove the existence, given $u,v\in\hh$, we define a linear form
on $\ccgg$ in the following way.  
Let $\beta\in\ccgg$, $K\subset G\times G$ be its support, $K_1$ and $K_2$ the
projection of $K$ on the first and second space, respectively. Fixed $g\in G$,
the map $g'\mapsto \beta(g',g)=:\beta_g$ is in $\ccg$, so the function 
$$G\ni g\mapsto \psi(g):=\scal{M(\beta_g)\pi(g)u}{v}\in\cuno$$
is well defined. We claim that $\psi\in\ccg$  Indeed, given $g_1,g_2\in G$ 
\mfi
|\psi(g_1)-\psi(g_2)| & = & |\scal{M(\beta_{g_1})\pi(g_1)u}{v} - \scal{M(\beta_{g_2})\pi(g_2)u}{v} | \\
& \leq &  |\scal{M(\beta_{g_1})(\pi(g_1)-\pi(g_2)u)}{v} | +
| \scal{M(\beta_{g_2}-\beta_{g_1})\pi(g_2)u}{v} |\\
(\mathrm{Eq.}\eqn{19})& \leq &  4 C_{K_1}\nor{v}\nor{E(X)}  \left(\sup_{G\times G}|\beta(g',g)|
  \nor{\pi(g_1)-\pi(g_2)u} \right. \\
&  & + \left. \sup_{g'\in G}|\beta(g',g_1) -\beta(g',g_2)| \nor{u} \right),
\mff
since  $\pi$ and  $\beta$ are continuous, also $\psi$ is continuous.  By
Eq.\eqn{19}, one has that, for all $g\in G$
$$|\psi(g)|  \leq   4  C_{K_1} \sup_{g'\in G}|\beta(g',g)| \nor{u}\nor{v}
\nor{E(X)},$$
so $\supp\psi\subset K_2$ and $\psi\in\ccg$. 

It follows that there is an operator $\Lambda(\beta)$ in $\hh$ such that 
\mfni
\scal{\Lambda(\beta) u}{v} & = & \intg{\psi(g)}{g}=
\intg{\scal{M(\beta_g)\pi(g)u}{v}}{g}\nonumber \\  
\nor{\Lambda(\beta)} & \leq &  C_K \sup_{G\times G}|\beta(g',g)|  \nor{E(X)}\label{1},
\mfnf
where $C_K=4 C_{K_1}\ \mug(K_2)$ depends only on the support of $\beta$. In
particular one has that, if $f_1,f_2\in\ccg$,
\mfni
\scal{\Lambda(f_1\otimes  f_2)u}{v} & = & \intg{f_2(g)\scal{M(f_1)\pi(g)u}{v}}{g}\nonumber \\
 & = & \scal{M(f_1)\pi(f_2)u}{v},\label{3}
\mfnf
and, for all $h\in H$ and $\beta\in\ccgg$,
\mfni
\scal{\Lambda(\beta(\cdot\,h,\cdot))u}{v} & = &
\intg{\scal{M(\beta_g(\cdot\,h))\pi(g)u}{v}}{g} \nonumber \\
\text{(Eq.\eqn{6})} & = & \deh(h\uno) \intg{\scal{M(\beta_g(\cdot))\pi(g)u}{v}}{g}
\nonumber\\
& = & \deh(h\uno)\scal{\Lambda(\beta(\cdot,\cdot))u}{v}.\label{7}
\mfnf

Fixed $u,v\in\hh$, by Eq.\eqn{1}, it follows that the linear form $\beta\mapsto
\scal{\Lambda(\beta) u}{v}$ is continuous with respect to the natural topology
of $\ccgg$, there is a measurable complex function
$\eta_{u,v}$ of modulo $1$ and a Radon measure $\lambda_{u,v}$ on $G\times G$ such that
$$\scal{\Lambda(\beta) u}{v} = \int_{G\times G} \beta(g,g')\eta_{u,v}(g,g')\,
d\lambda_{u,v}(g,g'),$$
see, for example, \cite[Ch. XIII, Sec.~16]{dieu3},

Given $f_1,f_2\in\ccg$ and $u,v\in\hh$, then, for all $f\in\ccg$
\mfi
\scal{M(f)\pi(f_1)u}{\pi(f_2)v} & = & \intg{\overline{f_2(g)}
  \scal{M(f)\pi(f_1)u}{\pi(g)v}}{g} \\
\text{(Eq.\eqn{2})} & = & \intg{\overline{f_2(g)} \scal{M(f^{g\uno})\pi(g\uno)\pi(f_1)u}{v}}{g} \\
\text{(Eqs.\eqn{0}, \eqn{3})} & = & \intg{\overline{f_2(g)}
  \scal{\Lambda(f^{g\uno}\otimes f_1^{g\uno})u}{v}}{g} \\
& = &   \intg{\overline{f_2(g)}\int_{G\times G}f(gg_1)f_1(gg_2)
  \eta_{u,v}(g_1,g_2)d\lambda_{u,v}(g_1,g_2)}{g} \\
& & \int_{G\times G\times G}f(gg_1)f_1(gg_2)\overline{f_2(g)}
  \eta_{u,v}(g_1,g_2)d\mug(g)d\lambda_{u,v}(g_1,g_2) \\
(g\mapsto gg_1\uno ) & = &  \int_{G\times G\times G}f(g)f_1(gg_1\uno
g_2)\overline{f_2(gg_1\uno)} \ \times\\
& & \room 
 \dgg(g_1\uno) \eta_{u,v}(g_1,g_2)d\mug(g)d\lambda_{u,v}(g_1,g_2) \\
& = & \intg{f(g)\phi_{\pi(f_1)u,\pi(f_2)v}}{g}, 
\mff 
where, for all $g\in G$,
\mfi
\phi_{\pi(f_1)u,\pi(f_2)v}(g) & = & \int_{G\times G}f_1(gg_1\uno g_2)\overline{f_2(gg_1\uno)}\dgg(g_1\uno)
  \eta_{u,v}(g_1,g_2)d\lambda_{u,v}(g_1,g_2),
\mff
which is a continuous function being $f_1,f_2\in\ccg$. By linearity, it
follows that, for all $u,v\in\gdo_\pi$ and $f\in\ccg$, there is a continuous
function $\phi_{u,v}$ such that 
$$\scal{M(f)u}{v} = \intg{f(g)\phi_{u,v}}{g}.$$
\vspace{.3cm}

\noindent{\em Step }3). We construct a Hilbert space $\kk$, which will carry the
representation $\sigma_E$.
 
Let $\phi$ be the sequilinear form defined on $\gdo_\pi\times\gdo_\pi$ as
$ \phi(u,v)  =  \phi_{u,v}(e)$. Clearly, if $f_1,f_2\in\ccg$ and $u,v\in\hh$
\mfni 
\phi(\pi(f_1)u,\pi(f_2)v) & = & \int_{G\times G}f_1(g_1\uno
g_2)\overline{f_2(g_1\uno)} \dgg(g_1\uno) \eta_{u,v}(g_1,g_2)d\lambda_{u,v}(g_1,g_2)\nonumber \\
& = & \scal{\Lambda(f_1\bullet f_2)u}{v} \label{5}
\mfnf 
with $(f_1\bullet f_2)(g_1,g_2) =f_1(g_1\uno
g_2)\overline{f_2(g_1\uno)}\dgg(g_1\uno)$. By Eqs.\eqn{0}, ~\eqn{5} and
dominated convergence theorem, the map
$$(g,g')\mapsto \phi(\pi(g)\pi(f_1)u,\pi(g')\pi(f_2)v)$$
is continuous on $G\times G$ and
$$\phi_{\pi(f_1)u,\pi(f_2)v}(g) = \phi(\pi(g\uno)\pi(f_1)u,\pi(g\uno)\pi(f_2)v).$$
By linearity, it follows that, for all $u,v\in\gdo_\pi$
\beeq{10}{(g,g')\mapsto \phi(\pi(g)u,\pi(g')v)\text{\ is continuous}}
and, for all $f\in\ccg$,
\beeq{11}{\intg{f(g)\phi(\pi(g\uno)u,\pi(g\uno)v)}{g} =
  \intg{f(g)\phi_{u,v}(g)}{g} = \scal{M(f)u}{v}.}
The form $\phi$ is non-negative, since, by construction, 
\mfi
\scal{M(f)u}{u} & = & \intg{f(g)\phi_{u,u}(g)}{g} \\
                & = & \int_{X} \left(\inth{f(gh)}{h}\right)
                d\scal{E(p(g))u}{u}
\mff
which is clearly non-negative, then $\phi_{u,u}(g)\geq 0$ $\mug$-almost
everywhere and, since $\phi_{u,u}$ is continuous, $\phi_{u,u}(e)\geq 0$. 

Let $\kk$ be the closure of the quotient space of $\gdo_\pi$ over the kernel
of $\phi$ with respect to scalar product induced by $\phi$ and $A$ be the
map from $\gdo_\pi$ to $\kk$ mapping $v\in\gdo_\pi$ into its equivalence class
$Av$, viewed in a natural way as an element of $\kk$.
 
We claim that $\kk$ is separable (so that $\kk$ is in fact a Hilbert space). 
Since $\nn:=A\gdo_\pi$ is dense in $\kk$, it is sufficient to show that $\nn$
is separable. Since $G$ is second countable, there is a denumerable
family $\set{f_n}_{n\in\mathbb N}$ in $\ccg$ such that for any $f\in\ccg$ and
$\eps>0$ there is a compact set $K$ and $f_n$ satisfying
\mfni
\supp (f-f_n) & \subset & K \label{4}\\
\sup_{g\in G} |f(g)-f_n(g)| & < & \eps\nonumber.
\mfnf
Moreover, since $\hh$ is separable, there is  a denumerable
family $\set{u_m}_{m\in\mathbb N}$ dense in $\hh$. 
We claim that $\set{A\pi(f_n)u_m}_{n,m\in\mathbb N}$ is dense in $\nn$.

Indeed, given $f\in\ccg$ and $u\in\hh$, let $f_n$ and $u_m$ such that
Eq.\eqn{4} holds, $\nor{u-u_m}<\eps$ and $\nor{u_m}\leq 2\nor{u}$. Then
\mfi 
\nor{A\pi(f)u-A\pi(f_n)u_m}_\kk & \leq & \nor{A\pi(f-f_n)u_m}_\kk +
\nor{A\pi(f)(u-u_m)}_\kk \\
& = & \sqrt{\phi_{\pi(l)u_m,\pi(l)u_m}}+
\sqrt{\phi_{\pi(f)v,\pi(f)v}}, 
\mff
where $l=f-f_n$ and $v=u-u_m $. Then, using Eq.\eqn{1} and~\eqn{5},
\mfi
\phi_{\pi(l)u_m,\pi(l)u_m} & = & \scal{\Lambda(l\bullet l)u_m}{u_m}\\
& \leq & C_K  \sup_{G\times G}|l(g_1\uno
g_2)\overline{l(g_1\uno)}\dgg(g_1\uno)|  \nor{E(X)} \nor{u_m}^2\\
& \leq & C_K \sup_{g\in K}(\dgg(g\uno)) (\sup_{g\in G} |f(g)-f_n(g)|)^2
\nor{E(X)} \nor{u_m}^2 \\
& \leq &  4 C_K \sup_{g\in K}(\dgg(g\uno))\nor{E(X)}\nor{u}^2  \eps^2, 
\mff
where $C_K$ depends only on $K$. In the same way,
\mfi
\phi_{\pi(f){v},\pi(f){v}} & = & \scal{\Lambda(f\bullet f)v}{v} \\
& \leq
 & C_K  \sup_{G\times G}|f(g_1\uno
g_2)\overline{f(g_1\uno)}\dgg(g_1\uno)|  \nor{E(X)} \nor{u-u_m}^2 \\
& \leq &   C_K \sup_{g\in K}(\dgg(g\uno))\nor{E(X)} (\sup_{g\in G} |f(g)|)^2 \eps^2.
\mff
From the above inequalities, one has that
$$ \nor{A\pi(f)u-A\pi(f_n)u_m}_\kk  \leq C' \eps,$$
where $C'$ is a suitable constant depending only on $f$ and $u$. 
Since the set $\set{A\pi(f)u}$ spans $A\gdo_\pi$, the claim follows. 
\vspace{.3cm}

\noindent{\em Step }4). We define a representation $\sigma_E$, denoted in
the following simply by $\sigma$, and an isometry $W$ satisfying
Eqs.\eqn{uno},\eqn{due} and\eqn{tre}.

To this aim, we first prove that, for all $h\in H$ and $u,v\in\gdo_\pi$
\beeq{12}{\phi(\pi(h)u,\phi(h)v)=\rap{h\uno} \phi(u,v).}
We can always assume that $u=v=\pi(f)w$ for some $f\in\ccg$ and
$w\in\hh$. Then, by Eq.\eqn{0}, 
\mfi
\phi(\pi(h)\pi(f)w,\pi(h)\pi(f)w) & = & \phi (\pi(f^h)w,\pi(f^h)w) \\
 & = & \scal{\Lambda(f^h\bullet f^h)w}{w} \\
 & = & \dgg(h)\scal{\Lambda((f\bullet f)(\cdot\,h,\cdot))w}{w} \\
\text{(Eq.\eqn{7})} & = & \rap{h\uno} \scal{\Lambda((f\bullet f)(\cdot,\cdot))w}{w} \\
 & = & \rap{h\uno} \phi(\pi(f)w,\pi(f)w).
\mff
From Eq.\eqn{12}, it follows that there is an isometric operator $\sigma_h$ in $\kk$
such that, for all $h\in H$, 
\beeq{8}{\sigma_h A u = \sqrt{\rap{h}} A\pi_h u \room u\in\gdo_\pi.}
We claim that $h\mapsto \sigma$ is a representation of $H$. The algebraic properties
are clear and, since $\sigma_h$ is isometric for all $h\in H$, it is
sufficient to show that, given $u,v\in\gdo_\pi$, the map  
$$h\mapsto \scal{\sigma_h Au}{Av}_\kk= \sqrt{\rap{h}} \phi(\pi_h u, v)$$ 
is continuous and this fact follows from Eq.\eqn{10}. We denote by $\sigma$
the representation defined by Eq.\eqn{8}.  

Moreover, we claim that $A$ is an admissible map with respect to $(\sigma,H)$. 
By construction, Eq.\eqn{semi} is satisfied. Moreover, for all $u,v\in\gdo_\pi$, the
map $g\mapsto \scal{A\pi(g\uno)u}{Av}_\kk$ is continuous by  Eq.\eqn{10} and, hence,  
$g\mapsto A\pi(g\uno)u$ is $\mug$-measurable.
Let now $f\in\ccg$ non-negative and $u\in\gdo_\pi$, by Eq.\eqn{11}
\mfi
\scal{M(f)u}{u} & = & \intg{f(g)\phi(\pi(g\uno)u,\pi(g\uno)u)}{g} \\
                & = & \intg{f(g)\nor{A\pi(g\uno)u}^2} {g}\\
&  = &   \intg{f(g)\nor{A\pi(g\uno)u}^2\inth{\theta(gh)}{h}}{g} \\
(g\mapsto gh\uno) & = &   \inth{\intg{\dgg(h\uno)
f(gh\uno)\nor{A\pi(hg\uno)u}^2 \theta(g)}{g}}{h} \\
(\text{Eq.\eqn{semi}}, h\mapsto h\uno) & = & 
\intg{\left(\inth{f(gh)}{h}\right)\nor{A\pi(g\uno)u}^2  \theta(g)}{g}.
\mff
By definition of $M(f)$ and  with notation of Lemma~\ref{tilde}, one has
that
\beeq{20}{\int_X \ft(x) d\scal{E(x)u}{u} =\intg{\ft(p(g))\nor{A\pi(g\uno)u}^2  \theta(g)}{g}.}
By Lemma~\ref{tilde}, there is a sequence $(f_n)_{n\in\nat}$ in $\ccg$
positive such that $(\ft_n)_n$ is a partition of the unit of $X$. Then
\mfi
\scal{E(X)u}{u} & = & \sum_n\int_X \ft_n(x) d\scal{E(x)u}{u} \\
 \text{(Eq.\eqn{20})}& = &  \sum_n \intg{\ft_n(p(g))\nor{A\pi(g\uno)u}^2 \theta(g)}{g}.
\mff
By monotone convergence theorem, the map $g\mapsto
\nor{A\pi(g\uno)u}^2 \theta(g)$ is $\mug$-integrable and
$$ \scal{E(X)u}{u} = \intg{\nor{A\pi(g\uno)u}^2 \theta(g)}{g}.$$
This shows that $A$ is an admissible map. 

Let $W_A$ be the corresponding wavelet operator. From the above equation, one
has that, for all $u\in\gdo_\pi$, $\scal{W_Au}{W_Au}=\scal{E(X)u}{u}$, that
is, by density, $W_A^*W_A=E(X)$. By Weak Frobenius theorem, $W_A$ intertwines
$\pi$ with $\ind$ and, by definition of localisation observable,  $E(X)$ is
injective. Then, by polar decomposition, there is an isometry $W$ such that
$W_A= WE(X)^{\frac 12}.$  Since $W$ intertwines $\pi$ with $\ind$, $W$
satisfies Eq.\eqn{uno}. 

To prove Eq.\eqn{due}, let $f\in\ccg$ and $u\in\gdo_\pi$,
using the definition of $\lind$, 
\mfi
\int_X\ft(x) d\scal{W_A^* \lind(x)W_A u}{u} & = & \intg{\ft(p(g))
    \scal{(W_Au)(g)}{(W_Au)(g)}_\kk \theta(g)}{g} \\
& = & \intg{\ft(p(g))\nor{A\pi(g\uno)u}^2  \theta(g)}{g} \\
\text{(Eq.\eqn{20})} & = & \int_X \ft(x) d\scal{E(x)u}{u}.
\mff
By Riesz-Markov theorem and the surjectivity of the map $f\mapsto \ft$, see
Lemma~\ref{tilde}, it follows that, for all $Y\in\boxx$ and $u\in\gdo_\pi$,
$$\scal{E(Y)u}{u} =\scal{ W^*_A \lind(Y)W_A u}{u}.$$
By density and the definition of $W$ it follows the claim.

For Eq.\eqn{tre}, it is sufficient to prove that the closed subspace
$$\mm:=\set{ F\in\hind\ | \ \scal{F}{\lind(Y)W_Av}=0, \forall  Y\in\boxx,
  v\in\hh}$$
is the null space. Using Eq.\eqn{impri} and the fact that $W_A$ commutes with the action of $G$,
it follows that $\mm$ is a $G$-invariant closed subspace of $\hind$. In
particular, if $f\in\ccg$ and $F'\in\mm$, $F=\ind(f)F'$ is in $\mm$ and, due to
Prop.~\ref{ind}, $F$ is a continuous function. Let now $Y\in\boxx$ and $v\in\gdo_\pi$, then
\mfni
0 & = & \scal{F}{\lind(Y)W_Av} \nonumber\\
 & = & \int_{p\uno(Y)} \scal{F(g)}{(W_Av)(g)}_\kk \theta(g)\ d\mug(g) \label{14}.
\mfnf
Since $v\in\gdo_\pi$, by Weak Frobenius theorem, $W_Av$ is continuous and, hence,
also the map $\scal{F(g)}{(W_Av)(g)}_\kk$ is continuous. Due to this and the
fact that $F,W_Au\in\hind$, one has that for all $h\in H$ and {\em for all}
$g\in G$
\beeq{13}{\scal{F(gh)}{(W_Av)(gh)}_\kk=\rap{h}\scal{F(g)}{(W_Av)(g)}_\kk.}
Let 
$$Y'=\set{g\in G \ | \ \scal{F(g)}{(W_Av)(g)}_\kk \leq 0 },$$
which is closed. Since $\rap{h}$ is strictly positive, due to Eq.\eqn{13}, for all $h\in H$,
$Y'h=Y'$. Defined $Y=p(Y')$, which is closed since $p$ is an open map, one has
that $Y'=p\uno(Y)$ and, using Eq.\eqn{14}, it follows that
$\scal{F(g)}{(W_Av)(g)}_\kk =0$ $ \theta\mug$-almost all $g\in Y'$. By
Lemma~\ref{theta} ($Y'h=Y'$) and the continuity, the above equality holds for
all $g\in Y'$. Repeating the above argument, one
concludes that $\scal{F(g)}{(W_Av)(g)}_\kk =0$ for all $g\in G$. 
By definition of $W_A$, $\scal{F(g)}{A\pi(g\uno)v}_\kk=0$ and,
since $A\gdo_\pi$ is dense in $\kk$ and $\pi(g\uno)$ is unitary, it follows
that $F(g)=0$ for all $g\in G$, that is $F=0$. Since $\set{\pi(f)F'\ | \ f\in\ccg, F'\in\mm}$
is dense in $\mm$, one has that $\mm=0$. The claim is now clear. 
\vspace{0.3cm}

\noindent{\em Step }5). We show that $\sigma$ is unique up to an equivalence. 

Let $\tau$ be another representation of $H$ acting in $\kk'$ and $W'$ the isometry from
$\hh$ to $L^\tau$ such that Eqs.\eqn{uno}, \eqn{due} and \eqn{tre} hold. By
Weak Frobenius theorem, there is an admissible map $B$ from $\pi$ with respect
to $(H,\tau)$ such that $W_B=WE(X)^{\frac 12}$. Given $f\in\ccg$
and $u\in\gdo_\pi$, applying Eq.\eqn{due} with $W'$ instead of $W$, one has that
\mfi
\scal{M(f)u}{u} & = & \int_X \ft(x)\ d\scal{E^\tau(x)W_Bu} {W_Bu} \\
& = & \intg{\ft(p(g))\nor{B\pi(g\uno)u}^2\theta(g)}{g} \\
& = & \inth{\intg{f(gh)\nor{B\pi(g\uno)u}^2\theta(g)}{g}}{h}\\
(g\mapsto gh\uno,h\mapsto h\uno) & = & 
 \inth{\intg{f(g)\nor{B\pi(g\uno)u}^2 \theta(gh)}{g}}{h}\\
& = & \intg{f(g)\nor{B\pi(g\uno)u}^2}{g}.   
\mff
On the other hand, due to Eq.\eqn{11},
$$\intg{f(g)\nor{A\pi(g\uno)u}^2 }{g} =  \intg{f(g)\nor{B\pi(g\uno)u}^2}{g}.$$
By usual arguments, it follows that $\nor{Au}_\kk=\nor{Bu}_{\kk'}$. Hence, there is a unique
isometric operator $t$ from $\kk$ to $\kk'$ such that
$$tAu=Bu\room u\in\gdo_\pi.$$
By definition of $\sigma$ and the fact that $B$ satisfies
Eq.\eqn{semi}, $t$ intertwines $\sigma$ and $\tau$. Since Eq.\eqn{tre} holds
both for $W$ and $W'$, one has that $t$ is in fact unitary. This shows that
$\tau$ is equivalent to $\sigma$. 
\vspace{0.3cm}

\noindent{\em Step }6).  We characterise the condition that $E$ is projective.

Assume now that $WE(X)^{\frac 12}$ is unitary, then since $\lind$ is projective,
see Example \ref{ex}, and Eq.\eqn{due}, it follows that also $E$ is
projective. Conversely, assume that $E$ is projective. Since $E$ is
normalised, $E(X)^{\frac 12}=I$ and $W=W_A$. For all $Y,Z\in\boxx$ and $u,v\in\hh$,
\mfi
\scal{WW^*\lind(Y)W u}{\lind(Z)W v}  & = & \scal{W E(Y) u}{\lind(Z)W v}\\
& = & \scal{W^* \lind(Z) W E(Y) u}{ v} \\
& = &  \scal{E(Z\cap Y) u}{ v} \\
& = & \scal{W^* \lind(Z\cap Y) W E(Y) u}{ v} \\
& = & \scal{\lind(Y)W u}{\lind(Z)W v},
\mff
since both $E$ and $\lind$ are projective. By Eq.\eqn{tre}, $\set{\lind(Y)W u}$
is total in $\hind$ so that $WW^* =I$ and, since $W$ is an isometry, $W$ is unitary.
\vspace{0.3cm}

\noindent{\em Step }7). We show that the equivalence class of $E$ defines
the equivalence class of $\sigma_E$.  

Let  $E'$ as in the statement of the theorem and $T$ a unitary operator such that
\mfni
E(Y)'T & = & TE(Y)\room Y\in\boxx \label{17}\\
\pi(g)'T & = & T\pi(g) \room g\in G \label{18},
\mfnf 
where, here and in the following, we denote by a prime the the
objects that refer to $E'$. Given $f\in\ccg$, from Eq.\eqn{18}, if follows
that, $M'(f)T=TM(f)$. Let now $u\in\gdo_\pi$ and $v\in\gdo_{\pi'}$, by Lemma~\ref{garding} $Tu\in\gdo_{\pi'}$ and, by
Eq.\eqn{11} applied first to $E'$ and, then, to $E$, 
\mfi
  \intg{f(g)\phi'(\pi'(g\uno)Tu,\pi'(g\uno)v)}{g} &  = &\scal{M(f)'Tu}{v} \\
& = & \scal{M(f)u}{T^*v} \\
& = &  \intg{f(g)\phi(\pi(g\uno)u,\pi(g\uno)T^*v)}{g}. 
\mff
By standard arguments, one has that $\phi'(Tu,v) = \phi(u,T^*v).$
Since $T$ is unitary, for all $u,v\in\gdo_\pi$,
$$ \phi'(Tu,Tv) = \phi(u,v).$$
It follows that there is an unitary operator $t$ from $\kk$ to $\kk'$ such
that, for all $u\in\gdo_\pi$,
$$ tAu=A't u.$$
By definition of $\sigma$ and $\sigma'$ and Eq.\eqn{18}, one has
that, for all $h\in H$ and $u\in\gdo_\pi$,
$$\sigma'(h)tu=t\sigma(h)u.$$
Hence, by density, $\sigma'$ is equivalent to $\sigma$.
\end{proof}
We add some comments. With the notation of Example~\ref{ex} and
$\sigma=\sigma_E$, Eq.\eqn{due} implies that  $E$ is equivalent to a
localisation observable of the form $\lind_T$ where $T=WE(X)^{\frac 12}W^*$ is
a positive operator on $\hind$.  In particular, there
exists always a {\em normalised} localisation observable $E_0$ covariant with respect to
$\pi$ and acting in $\hh$ such that
\beeq{23}{E(Y)=E(X)^{\frac 12}E_0(Y)E(X)^{\frac 12}\room Y\in\boxx.}
If $E$ is normalised, Eq.\eqn{tre} 
implies that $(\hind,\lind)$ is the Neumark dilation of $(\hh,E)$. Moreover,
since our proof is independent on Mackey Imprimitivity 
theorem, it contains the Mackey's result as a particular
case and one can easily show that two
{\em projective} localisation observables $E$ and $E'$ are equivalent, if and only if $\sigma_E$
and $\sigma_{E'}$ are equivalent. However, if $E$ and $E'$ are not projective one has
only the {\em only if} part, as showed by the following example. 
\bg{example} Let $G=\mathcal T$ be the one dimensional torus and
$H=\set{e}$. Denoted by  $(f_1,f_2)$ the canonical basis of $\hh:=\cuno^2$,
define, for all $z\in {\mathcal T}$ and $Y\in\mathcal{B}({\mathcal T})$, 
\mfi
\pi(z) & = & z \scal{\cdot}{f_1} f_1 + z^2 \scal{\cdot}{f_2} f_2 \\
E(Y) & = & \mu(Y) \scal{\cdot}{f_1} f_1 + \mu(Y) \scal{\cdot}{f_2} f_2 \\
E'(Y) & =  & E(Y) + \int_Y \overline{z}\ d\mu(z) \scal{\cdot}{f_1} f_2 + \int_Y
z\ d\mu(z) \scal{\cdot}{f_2} f_1, 
\mff
where $\mu$ is the normalised Haar measure on ${\mathcal T}$.
By direct computation, one can check that $E$ is a projective localisation
observable and $E'$ is a normalised non-projective one, both covariant with
respect to $\pi$. Moreover, one has that $\sigma_E=\sigma_{E'}$ is the two
dimensional trivial representation of the identity, however $E$ and $E'$ are
not equivalent.
\end{example}
By Eq.\eqn{23}, one can always assume that localisation
observables are normalised. The following corollary  settles a correspondence between normalised localisation
observables $E$ and admissible maps $A$ satisfying Eq.\eqn{tight}, compare
with \cite{h}, \cite{mo}, \cite{scutaru}.
\bg{corollary}\noindent
Let $\pi$ be a representation of $G$ and $\theta$ as in Lemma~$\ref{theta}$.  
Given a normalised localisation observable $E$ covariant with respect to $\pi$,
there is an admissible map $A$ for $\pi$ with respect to $(H,\sigma_E)$ that
satisfies Eq.$\eqn{tight}$ and, for all $u,v\in\gdo_\pi$ and $Y\in\boxx$,
\beeq{22}{\scal{E(Y)u}{v}=\int_{p\uno(Y)}
\scal{A\pi(g\uno)u}{A\pi(g\uno)v}\theta(g)d\mug(g).}
In particular $\pi$  is square integrable modulo $(H,\sigma_E)$. 

Conversely, if $\pi$ is square integrable modulo $(H,\sigma)$ for some
representation $\sigma$  of $H$ and $A$ is an admissible map  for $\pi$ with
respect to $(H,\sigma)$ such that Eq.$\eqn{tight}$ holds, then Eq.$\eqn{22}$ defines 
a normalised localisation observable  covariant with respect to $\pi$.
\end{corollary}
\bg{proof}
For the first part, due to Generalised Mackey theorem, there is an isometry  $W$ such that 
Eqs.\eqn{uno} and~\eqn{due} hold. In particular, $\pi$ is square-integrable.
Since $W$ is an intertwining isometric operator, by Weak Frobenius
theorem, there is an admissible map $A$ such that Eq.$\eqn{tight}$ holds. 
By definition of $E^{\sigma_E}$, Eq.\eqn{22} follows. 

Conversely, assume that $\pi$ is square integrable modulo $(H,\sigma)$ for
some $\sigma$. By Corollary~\ref{def}, there are admissible maps $A$
satisfying Eq.\eqn{tight}. In particular the corresponding wavelet operator $W_A$ is
isometric and intertwines $\pi$ and $\ind$. Then, the map
$$Y\mapsto   W_A^* \lind(Y)W_A=:E(Y)$$
is a normalised localisation observable covariant with respect to $\pi$, explicitely
given by Eq.\eqn{22}.    
\end{proof}

\end{document}